# Penetration Testing: A Roadmap to Network Security

Mr. Nitin A. Naik, Mr. Gajanan D. Kurundkar, Dr. Santosh D. Khamitkar, Dr. Namdeo V. Kalyankar

**Abstract:** Network penetration testing identifies the exploits and vulnerabilities those exist within computer network infrastructure and help to confirm the security measures. The objective of this paper is to explain methodology and methos behind penetration testing and illustrate remedies over it, which will provide substantial value for network security Penetration testing should model real world attacks as closely as possible. An authorized and scheduled penetration testing will probably detected by IDS (Intrusion Detection System). Network penetration testing is done by either or manual automated tools. Penetration test can gather evidence of vulnerability in the network. Successful testing provides indisputable evidence of the problem as well as starting point for prioritizing remediation. Penetration testing focuses on high severity vulnerabilities and there are no false positive.

**Index Terms**— Attacks, Intruder, Penetration Testing, Social Engineering attacks,

—————————— ♦ ——————————

## 1 INTRODUCTION

Penetration testing can reveal to what extent the security of IT systems is threatened by attacks by hackers, crackers, etc., and whether the security measures in place are currently capable of ensuring IT security. For a clearer picture of the risks to IT security, the term "penetration test" and the methods used for testing were established in 1995 when the first Unix-based vulnerability scanner "SATAN" was introduced. At that time the program was the first tool that was able to automatically scan computers to identify vulnerabilities. Nowadays, there are a number of freeware and commercial vulnerability scanners, most of which have an updatable database of known hardware and software vulnerabilities. These tools are a convenient way of identifying vulnerabilities in the systems being tested and therefore of determining the risks involved. Penetration testing is also reffered as Pen Testing or White Hat Attack because good guys are also try to break in to system.

## 2. INTRUDERS

The term "Hacker" is used to refer to any person who illegally logged into other IT systems without authorization. "Hackers" are regarded as being intelligent programmers who target security loopholes in IT systems for technical reasons they are not destroy anything but only for curiosity they enter into someone else's system. "Crackers" are people with criminal energy who utilize weak points of IT systems to gain illegal advantages, social attention or respect.[1] They are normally peoples who get access of complete software by cracking the serial or password of software. "Script kiddies" are usually intruders lacking in-depth background knowledge and driven by curiosity who mainly direct attack tools downloaded from the Internet against arbitrary or prominent targets.[1] Intruders can have a range of motives for carrying out attacks on IT infrastructure.[1]

## 3. MAJOR NETWORK ATTACKS

There are many ways of manipulating, Illigally updating or damaging IT Networks and of preparing an attack on IT Network.

### 3.1 Network-based attacks:

"Network-based attacks" use network protocol functionalities for exploitation and damage. Network-based attacks are extended for Port Scanning, IP Spoofing, Sniffing, Session Hijacking, DoS attacks, buffer overflow attack impairs the target system by overflowing a buffer whose boundry is unchecked.[2] ,format string attacks, and other exploitation of vulnerabilities in operating systems, application systems and network protocols.

### 3.2 Social engineering attacks:

Social engineering attacks are attempts to manipulate people with privileged knowledge to make them reveal security-related information such as passwords to the attacker. The ranges of possible attacks are wide with this technique. In its broadest sense, social engineering can also cover situations in which security related information is obtained by extortion. Social Engineering Penetration test works best when there are specific policies and procedures that are being tested.[3]

## 4. PENETRATION TESTING : STEPS

Following steps are followed for penetration testing over the network:

### 4.1 Information about the target system

Every Computer that can be accessed over the internet have an official IP address. Some organizations provides the information regarding the block of ip addresses about the systems over internet.

————————————————

- Mr. N.A.Naik is working as lecturer with Dept. of Computer Science and IT , Yeshwant College Nanded
- Mr. G.D.Kurundkar is working as Lecturer with Dept. of Computer Science S.G.B. College Purna
- Dr. S.D. Khamitkar is working as Director, School of Computational Sciences S.R.T.M.University , Nanded.
- Dr. N.V.Kalyankar is working as Principal, Yeshwant College Nanded.

### 4.2 Scan target systems for services on offer
An attempt is made to conduct a port scan of the Systems(s) being tested, open ports being indicative of the applications assigned to them. [2]

### 4.3 Identify systems and applications
The names and version of operating systems and applications in the target systems can be identified by "fingerprinting".

### 4.4 Researching Vulnerabilities
Information about vulnerabilities of specific operating systems and applications can be researched efficiently using the information gathered. Every operating system have some loophole in it so that the data or the information which is stored in the operating system may be attacked by an intruder , so the researchers should gather the initial information of operating system and then try to penetrate the system by applying some rules.

### 4.5 Exploiting vulnerabilities
Detected vulnerabilities can be used to obtain unauthorized access to the system or to prepare further attacks. The quality and value of a penetration test depends primarily on the extent to which the test caters to the client's personal situation, i.e. how much of the tester's time and resources are spent on detecting vulnerabilities related to the IT infrastructure and how creative the tester's approach is. This process cannot be covered in the general description above, which is why there are huge differences in the quality of penetration testing as a service.

## 5. HOW IT WORKS ?
The following introduces the five phases of a penetration test based on the steps given above. The individual phases take place successively.

**Phase 1: Introductory Preparation**; It is difficult to fulfill the client's expectations without good grounding, such as reaching a settlement on the objectives of the penetration test. At the start of a penetration test the client's objectives must be clarified with him and defined. The performance of a penetration test without taking full account of the relevant legal provisions could have cost under criminal or civil law. The tester must therefore ensure that the test procedures are not going to violate legal provisions or contractual Settlements. The failure of a testing could also lead to alternative demands. All details agreed to should be put in writing in the contract.

**Phase 2: Investigation :** After the testing decision , objectives, scope, procedures, emergency measures ,limitations have been defined , taking account of the legal and organizational aspects and other conditions, the tester can start gathering information on the target system. This phase is the passive penetration test. The aim is to get complete information of the system and get information of shortcomings in the system. Depending on the size of the network to be examined, the test steps may be extremely time-consuming. These long test steps are usually performed automatically, the time required for them still needs to be taken into account in the planning. Thus a penetration test can take several days.

**Phase 3: Analyzing information and risks:** A successful, clear and economically efficient procedure must analyze and assess the information gathered before the test steps for actively penetrating the system. The analysis must include the defined goals of the penetration test, the possible risks to the system and the time required for evaluating the possible security flaws for the succeeding active penetration attempts. The aims in phase 4 are then selected on the basis of this analysis. From the list of identified systems the tester may choose the systems which have known potential vulnerabilities due to their configuration or the identified applications/services or those about which the tester is particularly knowledgeable. In a penetration test for which the number of target systems has been clearly defined in phase 2, this selection means that the number of target systems for phase 4 is automatically reduced. The restrictions must be broadly documented and justified because addition to the desired improvement in efficiency, they also lead to a reduction in the informative value of the penetration test and the client needs to be made aware of this.

**Phase 4: Active intrusion attempts:** Finally, the selected systems are actively attacked. This phase involves the highest risk within a penetration test and should be performed with due care. However, only this phase reveals the extent to which the supposed vulnerabilities identified in the investigation phase present actual risks. This phase must be performed if a verification of potential vulnerabilities is required. For systems with very high availability or integrity requirements, the potential effects need to be carefully considered before performing critical test procedures, such as the utilization of buffer overflow exploits. In a white-box test, a patch may need to be installed on critical systems before performing the test to prevent system failure. The test will probably not be able to locate any vulnerability, but will document the security of the system. Unlike a hacking attack, however, the penetration test is not complete – it will be continued.

**Phase 5: Final analysis:** As well as the individual test steps, the final report should contain an evaluation of the vulnerabilities found in the form of potential risks and recommendations for eliminating the vulnerabilities and risks. The report must guarantee the transparency of the tests and the vulnerabilities it found during testing. The findings of risks for IT security should be discussed in detail with the client after the successful completion of the test procedures.

For a successful penetration test that meets the client's expectations, the clear definition of goals is mainly essential. If goals cannot be achived efficiently, the tester should notify the client in the preparation phase and pro-pose alternative procedures such as an IT audit or IT security consulting services. Client goals that can be attained by penetration testing can be divided into four categories:

1. Improving security of technical systems
2. Identifying vulnerabilities
3. Having IT security confirmed by an external agency
4. Improving security of infrastructure of organization and personal infrastructure

The result of an IT penetration test should not be only list of existing vulnerabilities but also suggest specific solutions for their elimination because A penetration test is an authorized, local attempt to "hack" into a system, to identify exploitable weaknesses, and to reveal what systems and data are at risk.[4]

## 6. APPLICATIONS OF PENETRATION TESTING

- Understand and reduce the impact, frequency, and sternness of security incidents.
- Meet compliance and regulatory requirements that require security assessments[5].
- Optimize and prioritize resources used to remediate vulnerabilities[5].
- Gain peace of mind about security safeguards, controls, and policies[5].
- Identifies vulnerabilities and risks in your networking infrastructure.
- Validates the effectiveness of current security safeguards.
- Quantifies the risk to internal systems and confidential information.
- Raises executive awareness of corporate liability.
- Provides detailed remediation steps to prevent network compromise.
- Validates the security of system upgrades.
- Protects the integrity of online assets.
- Helps to achieve and maintain compliance with federal and state regulations
- Using an automated product allows you to consistently test your network and easily integrate the practice with your overall security program. This means you'll have more confidence in the overall security of network[5]
- It will give information about the how much information is publicly available?.
- 

## 7. LIMITATIONS OF PENETRATION TESTING

Now a days attackers or hackers are becomes more smart and intelligent and the new sercurity related problems in IT sercurities are reported very rapidly.To make a system more sercure it is necessary to take bulk tests at a time. A new security loophole may mean that a successful attack could take place immediately after a penetration test has been completed.[2] It is possible that the new security hole which is not discovered during testing may result into attack[6]. As definition of Testing service can only find the vulnerabilities and cannot prove the absence of vulnerabilities, although the independent client test consistently show the quality of penetration test. As penetration test report shows the methodology which is used during test and various procedures used during penetration test a person who have some experience about the network security can guage the throughness of the test.

## 8. CONCLUSION

This paper gives information about the penetration testing , its methodologies and its application. Highlights how an experienced security consultant is necessary for the good penetration and role of him to give security system to the host machine by expecting the security attacks. The institutions / offices / companies where the network system is installed , it is necessary to deploy the security personnal who knows the possible modern security attacks and try to develop a mechanism to overcome these security attacks. For this it is necessary that the penetration system should be accrurately and scientifically created and executed. While documenting the test results of penetration system a scientific and procedural approach should be there.The penetration testing results should be taken frequently as there are day-to-day modifications in the attacks over network. As per the new modified attacks over network the penetratrion test should be modified as per attack (The security personnal should think as hacker).The organizations where the penetration system is deployed should give roadmap to the security personal for the security measures. As a measurement tool, penetration testing is most powerful when fully integrated into the development process in such a way that findings can help improve design, implementation, and deployment practices.

## AUTHORS

**Nitin A. Naik** Completed M.Sc. Computer Science from S.R.T.M. University in 1997 and from Dec. 1997 working as Lecturer in Dept. of Computer Science & IT , Yeshwant College Nanded.From Jan.

2007 working as Network Administrator for campus wide network of same college. He is also life member of Indian Science Congress, Kolkata (India).

**Gajanan D. Kurundkar** Completed M.Sc. Computer Science from S.R.T.M. University Nanded in 2000. He joined as lecturer in Dept. of Commputer Science from June 2001 to Jan. 2006 , Currently working as Lecturer in Dept of Computer Science at Sri Gurubuddhi Swami College Purna (India) from Jan.2006 to till date. He is also life member of Indian Science Congress, Kolkata (India)

**Santosh D. Khamitkar:** Completed M.Sc. Computer Science from B.A.M. University , Aurangabad. in 1994. in 1995 he joined as Lecturer in School of Computational Sciences in S.R.T.M.University Nanded. He Completed his Ph.D. from S.R.T.M. University in 2009 and currently he is working as Director , School of Computational Sciences , S.R.T.M. University , Nanded. He is Technical Advisor (Freelance) of Portal Infosys, Nanded. He is also Research Guide for Computer Studies in S.R.T.M. University , Nanded. He is life member of Indian Science Congress, Kolkata (India)

**Namdeo V. Kalyankar:** Completed M.Sc. Physics from B.A.M. University , Aurangabad. in 1980. in 1980 he joined as Lecturer in Department of Physics in yeshwant College,Nanded. In 1984 he completed his DHE. He Completed his Ph.D. from B.A.M.University in 1995. From 2003 he is working as Principal since 2003 to till date in Yeshwant college Nanded. He is also Research Guide for Computer Studies in S.R.T.M. University , Nanded. He is also worked on various bodies in S.R.T.M. University Nanded. He also published research papers in various international/ national journals. He is peer team member of NAAC (National Assessment and Accreditation Council)(India). He published a book entitled " DBMS Concept and programming in Foxpro". He also got "Best Principal" award from S.R.T.M. University, Nanded(India) in 2009. He is life member of Indian Science Congress , Kolkata (India). He is also honored with Fellowship of Linnean Society of London (F.L.S.) on 11$^{th}$ Nov. 2009